\documentclass[prb,twocolumn,english,superscriptaddress]{revtex4-1}
\usepackage{amsmath,amssymb,mathrsfs}

\usepackage{amsfonts,bm}
\usepackage{amsmath}
\usepackage{amssymb}
\usepackage{amsthm}
\usepackage{graphicx}
\usepackage{graphics}
\usepackage{subfigure}
\usepackage{color}
\usepackage{bm}
\usepackage{hyperref}
\usepackage{rotating}
\usepackage{comment}

\newcommand{\be}{\begin{equation} }
\newcommand{\ee}{\end{equation} }
\newcommand{\ba}{\begin{eqnarray} }
\newcommand{\ea}{\end{eqnarray} }

\newcommand{\bpm}{\begin{pmatrix}}
\newcommand{\epm}{\end{pmatrix}}
\newcommand{\bmm}{\begin{matrix}}
\newcommand{\emm}{\end{matrix}}

\newcommand{\la}{\label}
\newcommand{\p}{\partial}
\newcommand{\bea}{\begin{eqnarray}}
\newcommand{\eea}{\end{eqnarray}}

\makeatother

\usepackage{babel}
\begin{document}

\title{Odd viscosity in two-dimensional incompressible fluids}
 \author{Sriram Ganeshan}
 \affiliation{Simons Center for Geometry and Physics, Stony Brook, NY 11794, USA}
 \author{Alexander G.~Abanov}
\affiliation{Simons Center for Geometry and Physics, Stony Brook, NY 11794, USA}
\affiliation{Department of Physics and Astronomy, Stony Brook University, Stony Brook, NY 11794, USA}

\date{\today}

\begin{abstract}
In this work, we present observable consequences of parity violating odd viscosity term in incompressible 2+1D hydrodynamics. For boundary conditions depending on the velocity field (flow) alone we show that: (i) The fluid flow quantified by the velocity field is independent of odd viscosity, (ii) The force acting on a closed contour is independent of odd viscosity, and (iii) The odd viscosity part of torque on a closed contour is proportional to the rate of change of area enclosed by the contour with the proportionality constant being twice the odd viscosity. The last statement allows us to define a measurement protocol of {\it odd viscostance} in analogy to Hall resistance measurements. We also consider {\it no-stress} boundary conditions which explicitly depend on odd viscosity. A classic hydrodynamics problem with no-stress boundary conditions is that of a bubble in a planar Stokes flow. We solve this problem exactly for shear and hyperbolic flows and show that the steady-state shape of the bubble in the shear flow depends explicitly on the value of odd viscosity. 
\end{abstract}
\maketitle

\section{Introduction} 
Hydrodynamics is generally concerned with the classical fluid motion arising due to application of external forces~\cite{lamb1932hydrodynamics, landau1987fluid}. In spite of this simplistic premise, hydrodynamic framework is ubiquitously applicable to a vast range of physical phenomena ranging from sub-atomic to astronomical scales. Thus discovery of a new phenomenon in hydrodynamics often leads to far reaching implications to a wide class of systems. One such phenomenon was uncovered in the seminal work of Avron, Seiler and Zograf ~\cite{avron1995viscosity} where they showed that the viscosity of quantum Hall (QH) fluids at zero temperature is non-dissipative and is closely related to the adiabatic curvature on the space of flat background metrics. This non-dissipative viscosity is the anti-symmetric component of the total viscosity tensor (dubbed {\it odd viscosity}~\cite{avron1998odd}) is non-zero in the presence of broken time reversal or broken parity symmetry. The 2+1D is special since the odd viscosity term is compatible with isotropy. The general parity odd terms of the viscosity tensor in 3+1D  was previously considered in the context of plasma in magnetic field~\cite{landau1987lifshitz} and in hydrodynamic theories of superfluid He-3A \cite{helium-book}. 

The concept of odd viscosity in QH fluids was subsequently generalized to 2+1D hydrodynamics with dominant odd viscosity term in the stress. The generalized Navier-Stokes equations with odd viscosity may lead to counter intuitive effects. Avron~\cite{avron1998odd} showed examples of such effects e.g., the radial pressure on a rotating cylinder and chiral viscosity waves with quadratic dispersion in compressible fluids. The odd viscosity effects have been studied extensively in the context of QH fluids (where it is dubbed as Hall viscosity)~\cite{tokatly2006magnetoelasticity,tokatly2007lorentz,tokatly2007new,tokatly2009erratum,tokatly2009lorentz,haldane2009hall, read2009non,haldane2011geometrical,haldane2011self,hoyos2012hall, bradlyn2012kubo, yang2012band, hughes2013torsional, hoyos2014hall, can2014field,can2014fractional,can2015geometry,klevtsov2015geometric,klevtsov2015quantum, gromov2014density, gromov2015framing, gromov2016boundary}. However, realistic odd viscosity effects measurable in laboratory for general classical fluids with broken time reversal symmetries~\cite{wiegmann2014anomalous, lapa2014swimming, lucas2014phenomenology} have received less attention. The closest attempt to this end was made recently in Ref.~\onlinecite{lapa2014swimming} where the authors considered geometric theory of swimming~\cite{shapere1987self, shapere1989gauge, shapere1989geometry} in Stokes flow with odd viscosity. The torque acting on the surface of the swimmer depends explicitly on the odd viscosity and influences the swimming strokes. Most recently the effects of odd viscosity have also been studied in connection with dynamics of systems of active rotors\cite{banerjee2017odd}. For an elegant pedagogical introduction of odd viscosity in fluid dynamics, we refer readers to Refs.~\onlinecite{avron1998odd, lapa2014swimming}.

While the effects of odd viscosity are very prominent in compressible fluids\cite{avron1998odd}, they are more subtle in incompressible case. Our goal is to identify general observable effects of odd viscosity in incompressible 2+1D fluids and capture these effects in terms of a possible measurement protocol. Such a measurement protocol can potentially be adopted to the case of QH fluids. In this work we prove several exact statements on the observable consequences of odd viscosity in incompressible 2+1D fluids. We show that the observable effects sensitively depend on the type of boundary conditions imposed. The applicability of a particular set of boundary conditions to a particular fluid should be either analyzed starting from microscopic (kinetic) theory or determined experimentally. Here we consider two classes of boundary conditions -- (a) Flow dependent boundary conditions or more precisely {\it no-slip} boundary conditions where the layer of fluid in contact with a solid body has the same velocity as the body, and (b) {\it no-stress} or force matching conditions where the velocities of the surfaces bounding a fluid are not all specified, but the surface tractions acting on these boundaries are known. For the flow-dependent boundary conditions, we prove the following three exact statements: (i) The fluid flow quantified by the velocity field is independent of odd viscosity, (ii) The force acting on a closed contour is independent of odd viscosity, and (iii) The odd viscosity part of torque on a closed contour is proportional to the rate of change of area enclosed by the contour with odd viscosity being the proportionality constant. We emphasize that the above statements are not limited to the case of Stokes flow (cf.~Refs.~\cite{avron1998odd, lapa2014swimming}) and are valid for the most general constant density incompressible fluid in 2+1D. We further exploit statement (3) to define a physical observable which we coin as {\it odd viscostance} and define a measurement protocol to observe it. 

The above statements do not hold generically for no-stress or free surface boundary conditions. A classic example of free surface boundary in hydrodynamics is the problem of finding a steady state or a dynamical shape of two-dimensional bubbles in slow viscous flows (Stokes flow)~\cite{cox1963inviscid, richardson1968two, richardson1973two, tanveer1994bubble, crowdy1998theory}. We generalize the problem of bubble dynamics to include odd viscosity and show that the steady state shape of the bubble explicitly depends on odd viscosity.  

\section{Incompressible fluids with odd viscosity} 
Hydrodynamic equations for a 2+1D incompressible fluid with constant density ($\rho=1$) are the incompressibility condition and Navier-Stokes (NS) equation given by \cite{landau1987fluid}:
\begin{align}
	 \partial_i v_i=0\,,\qquad  D_t v_i=-\p_i p+\p_j \sigma_{ij}\,, 	
\end{align}
where $v_i$ ($i=1,2$) are components of the fluid velocity field and $D_t=\p_t+v_i\p_i$ denotes the material derivative.  For an isotropic fluid but with broken parity the viscous stress tensor can take the following general form \cite{avron1998odd} \footnote{Strictly speaking one might also add the antisymmetric part of the stress $\sim \epsilon_{ij}\omega$ \cite{banerjee2017odd} but we do not consider this possibility in this work.}:
\begin{align}
	\sigma_{ij}=\nu_e(\p_iv_j+\p_jv_i)+\nu_{o} (\p_{i}v_{j}^{*} +\p_{i}^{*}v_{j})\,.
 \la{sigmaij}
\end{align}
Here $\nu_e$ and $\nu_o$ are the shear viscosity and the odd viscosity, respectively. We use the notation $a_i^*=\epsilon_{ij}a_j$.
Using the incompressibility condition  and (\ref{sigmaij}) the NS equation can be rewritten as,
\begin{align}
	D_{t}v_i &= -\p_i \tilde p +\nu_e \Delta v_i \,.
 \label{eq:navierStokes}
\end{align}
The odd viscosity term has been absorbed by redefining pressure $\tilde p=p-\nu_o \omega$, where $\omega=\epsilon_{ij}\p_iv_j$ is the vorticity of the fluid. Taking curl of the above equation removes modified pressure out of the equation  and we obtain an exclusively flow dependent equation of motion written in terms of the vorticity $\omega$
\begin{align}
	D_{t}\omega = \nu_e \Delta \omega \,.
	\label{eq:vorteqn}
\end{align}
Eq.~\ref{eq:vorteqn} along with the incompressibility condition $\p_iv_i=0$ constitute two equations for two components of velocity. Pressure is not a ``state variable'' in incompressible hydrodynamics. It can be found from the NS equation (\ref{eq:navierStokes}) after finding the flow $\bm{v}(\bm{x},t)$ from (\ref{eq:vorteqn}) and the incompressibility condition.
{\it Prima facie} it may seem that fluid flow and the corresponding equation of state is completely independent of odd viscosity. However, we must recall that the fluid flow is specified by equations of motion in conjunction with  boundary conditions. Thus to locate the observable effects of odd viscosity we must analyze various boundary conditions and see how they dictate odd viscosity effects in the resulting dynamics. To this end we consider {\it no-slip} and  {\it no-stress} ({\it free surface}) boundary conditions. We first consider the odd viscosity effects for no-slip boundary conditions. 

\section{Exact results for no-slip boundary conditions} The no-slip boundary conditions enforce that the layer of fluid in contact with a solid body has the same velocity as the body
\begin{align}
	v_i|_{\text{surface}}=U_i\,.
 \label{eq:noslip}	
\end{align}
Assuming this condition to be true, we can make the following three statements.

{\bf Statement I (Flow):} {\it If boundary conditions of an incompressible flow depend only on the flow $\bm{v}(\bm{x},t)$ itself, the flow  does not depend on the value of odd viscosity $\nu_{o}$.} 

The above statement simply follows from Eqs.~\ref{eq:vorteqn} and  \ref{eq:noslip} which are completely independent of $\nu_o$. It has appeared already in Ref.~\cite{avron1998odd} and here we just make it more precise by explicitly specifying boundary conditions. As a corollary to the statement I, we note that changing $\nu_o\to-\nu_o$ would leave the flow unchanged (in QH fluids this reversal can be achieved by changing the direction of external magnetic field).

{\bf Statement II (Force):} {\it If boundary conditions of an incompressible flow depend only on the flow $\bm{v}(\bm{x},t)$ itself, the net force acting on a closed contour $\Gamma$ does not depend on the value of odd viscosity $\nu_{o}$.}

 The force applied by a fluid to a unit length of a contour is given by $f_j=n_i T_{ij}$, where $\bm{n}$ is a unit vector orthogonal to the contour outward to $\Gamma$ and  $T_{ij}$ is the stress tensor defined as $T_{ij} = -p\delta_{ij}+\sigma_{ij}$. Using the incompressibility condition we rewrite 
\begin{align}
	T_{ij}=-\tilde p \delta_{ij}+\nu_e(\partial_i v_j+\partial_j v_i)
	+2\nu_o \partial^*_i v_j\,.
\end{align}
Then the force $\bm{f}$ acting on a contour $\Gamma$ is given by:
\begin{align}
	f_j\Big|_\Gamma = -\tilde{p}n_j 
	+\nu_e n_i(\p_i v_j+\p_j v_i)- 2\nu_{o} n_i^*\p_{i}v_{j}\,
 \nonumber
\end{align}
or introducing the tangent direction $\bm{s}=-\bm{n}^*$ to the contour $\Gamma$
\begin{align}
	f_j\Big|_\Gamma = -\tilde{p}n_j  
	+\nu_e n_i(\p_i v_j+\p_j v_i)+2\nu_{o} \p_{\bm{s}}v_{j}\,.
	\label{eq:localforce}
\end{align}
In the above expression only the last term depends on $\nu_o$. Indeed by the statement I, the flow itself does not depend on odd viscosity and so is modified pressure $\tilde{p}$ which can be found from flow according to (\ref{eq:navierStokes}). We can then calculate the contribution from the odd viscosity term to the total force $F_j^o $ acting on the closed contour $\Gamma$ 
\begin{align}
	F_j^o = 2\nu_o\oint_\Gamma ds\,\p_{\bm{s}} v_j =0\,.
\end{align}
This completes the proof of Statement II. A corollary of this statement is the absence of any total lift force or Magnus force coming from the odd viscosity when no-slip boundary conditions are imposed. We emphasize that this statement holds in general moving and shape-changing contours and for the incompressible fluid of constant density. It is not limited to Stokes flow (cf. Refs.~\cite{avron1998odd, lapa2014swimming}). In particular, non-linear corrections (also known as Oseen's correction) to Stokes flow do not result in $\nu_o$ dependent lift force on a cylinder in contrast to recent claims~\cite{kogan2016lift}. We have shown that that there is no lift force due to odd viscosity. The correct definition of a lift force involves integration of the full momentum flux density tensor projected along the normal direction to the closed contour, not just an integration of the change of the pressure due to odd viscosity $p-\tilde p=\nu_o \omega$. The final form of the net force acting on a closed contour $\Gamma$ is given by,
 \begin{align}
		F_j=-\oint_{\Gamma} (\tilde p n_j+\nu_e \omega s_j-2\nu_e n_i \partial_i v_j) ds\,.
\end{align}
It can be determined once the flow $v_j$ is obtained. Notice that  the statement II is valid for flow dependent boundary conditions and only for the net force acting on a closed contour. Effects of $\nu_o$ are still observable if one measures local forces given by Eq.~(\ref{eq:localforce}).

{\bf Statement III (Torque):} {\it If boundary conditions of an incompressible flow depend only on the flow $\bm{v}(\bm{x},t)$ itself, the part of the net torque acting on a closed contour which depends on the value of odd viscosity $\nu_{o}$ is given by:
\begin{align}
	\mathcal{T}^o = 2\nu_o\oint_\Gamma ds\, v_{\bm{n}} = 2\nu^{o}\frac{d{\cal A}}{dt}\,,
 \la{oddtorque}
\end{align}
where ${\cal A}$ is the area enclosed by the contour.}

The net torque on a closed contour is given by $\mathcal{T}=\oint_\Gamma \tau\,ds$, where the local torque $\tau$ on a unit contour element can be written as,
\begin{align}
	\tau = \epsilon_{kj}x_{k}f_j\Big|_\Gamma 
	&= -\tilde{p}(x_{k}n_k^{*})+\nu_e n_i x_{k}(\p_i v_k^{*}+\p_k^{*} v_i)
 \nonumber \\ 
 	&+ 2\nu_{o} x_{k}\p_{\bm{s}}v_{k}^{*}  \,.
 \nonumber
\end{align}
The last term is the only $\nu_o$-dependent term and the total torque corresponding to this term on a closed contour is given by,
\begin{align}
	\mathcal{T}^o = 2\nu_o\oint_\Gamma ds\,x_{k}\p_{\bm{s}}v_{k}^{*}
	= 2\nu_o\oint_\Gamma ds\, v_{\bm{n}}=2\nu_{o}\frac{d{\cal A}}{dt} \,.
 \nonumber
\end{align}
The integral in the above equation is the rate of the change of the area ${\cal A}$ enclosed by the contour. The expression $\mathcal{T}^o =  2\nu_{o}\frac{d{\cal A}}{dt}$ has an obvious physical interpretation as a rate of the expulsion of intrinsic angular momentum from the area enclosed by the contour $\Gamma$. The odd viscosity is given by the half of the value of intrinsic angular momentum per particle $\nu_{o} = \frac{l}{2}$~\cite{read2009non, wiegmann2014anomalous, banerjee2017odd}. Pushing away this angular momentum results in the torque $\mathcal{T}^{o}= \frac{d ({\cal A}  l)}{dt}$. Eq.~\ref{oddtorque} has appeared first in  Ref.~\cite{lapa2014swimming} where it was derived for the swimming in the Stokes regime. The derivation presented in this section extends the validity of formula \ref{oddtorque} beyond the Stokes limit.

\textbf{\textit{Odd viscostance:}} Following statement III, we can construct an experimental protocol that can lead to the measurement of odd viscosity.  We define \emph{odd viscostance} as
\begin{align}
	\mathcal{N}^{o}\equiv   \frac{1}{2} \frac{\mathcal{T}}{d\mathcal{A}/dt}\,,
	\label{eq:oddvis}
\end{align}
where $\mathcal{T}$ is a net torque acting on an expanding circle with no-slip boundary conditions on the circle. In case of incompressible, uniform, isotropic, infinite 2D fluid with constant odd viscosity $\nu_{o}$ one finds $\mathcal{N}^{o}=\nu_{o}$. However, the quantity $\mathcal{N}^{o}$ characterizes the fluid system \emph{globally} and might change if defects and inhomogeneities are present in the fluid. Indeed, specializing to the case of the radially expanding circle  ($v_s=0$ at $\Gamma$)we obtain for the torque $\tau=2\nu_{o}v_{\bm n}+ \nu_e r\p_r v_{\bm s}$. The latter term drops out in the limit $\nu_e\to 0$ but is generally non-vanishing for finite $\nu_o$ and in the presence of inhomogeneities. The relation between $\mathcal{N}^{o}$ and $\nu_{o}$ is similar to the relation between Hall resistance and Hall resistivity. 

If one has an experimental ability to change the sign of $\nu_o$ the easiest way to extract the odd viscosity dependent part of the torque (\ref{oddtorque}) is to repeat the torque measurement for both $\nu_o$ and $-\nu_o$. Then $\mathcal{T}^o$ is given by $\mathcal{T}^o=(\mathcal{T}_{\nu_o}-\mathcal{T}_{-\nu_o})/2$ and the value of odd viscosity $\nu_o$ can be found from (\ref{oddtorque}). We emphasize that this protocol is robust against the presence of impurities (presence of other rigid obstacles) as long as they interact with the fluid via no-slip boundary conditions.

\section{No-stress boundary conditions} The above exact statements are limited to the case of no-slip boundary conditions. For the no stress or free surface boundary conditions the statement I (and therefore, II and III as well) would break down and may lead to a odd viscosity dependent flow. In the following we investigate few examples of incompressible fluids with odd viscosity and no-stress boundary conditions. 

\subsection{Expanding bubble}
Let us consider the simplest example of expanding bubble in an incompressible fluid with odd viscosity. The key equation in the no-stress condition for the case of inviscid bubble is given by,
\begin{align}
	n_iT_{ij}=0\,.
\end{align}
Here, we took the pressure inside the bubble as zero.
A particular ``stationary solution'' in polar coordinates $(r, \theta)$ satisfying Eq.~\ref{eq:vorteqn} along with the incompressibility condition is given by,
\begin{align}
	v_{r} = \frac{\gamma}{2\pi r}\,, \qquad v_{\theta} = \frac{\alpha \gamma}{2\pi r}\,,
\end{align}
where $\gamma = d\mathcal{A}/dt=2\pi R \dot R$ is a rate of the area change and $\alpha$ is some constant. For this particular solution $\omega=0$ everywhere outside the bubble. At the surface of the bubble we have for normal and tangent forces $f_{\mathbf{n}}=n_{j}f_{j}$, $f_{\mathbf{s}}=f_{j}s_{j}$ with $f_j$ given by Eq.~\ref{eq:localforce}. For the solution at hand the local forces along normal and tangential direction is given by, 
\begin{align}	
	f_{\mathbf{n}} &= -p+2\nu_e \p_{r}v_{r}+2\nu_{o}\frac{v_{\theta}}{r}\;\Big|_{r=R}\,,
 \\
 	f_{\mathbf{s}} &=\nu_e\left(\p_{r}v_{\theta}-\frac{v_{\theta}}{r}\right)
	+2\nu_{o}\frac{v_{r}}{r}\;\Big|_{r=R}\,.
\end{align}
The first equation defines the necessary air pressure inside the bubble. The no-stress condition for tangent component of the force
\begin{align} 
	f_{\mathbf{s}}\Big|_{r=R} =0
\end{align} 
gives $\alpha = \frac{\nu_{o}}{\nu_e}$, where the flow is then given by,
\begin{align}
	v_{r} = \frac{\gamma}{2\pi r}\,, 
	\qquad  v_{\theta} = \frac{\nu_{o}}{\nu_e} \frac{\gamma}{2\pi r}\,.
\end{align}
The tangent component of the flow $v_{\theta}$ explicitly depends and is entirely due to the odd viscosity $\nu_o$.
\begin{figure}[tb]
  \centering
\includegraphics[scale=0.32]{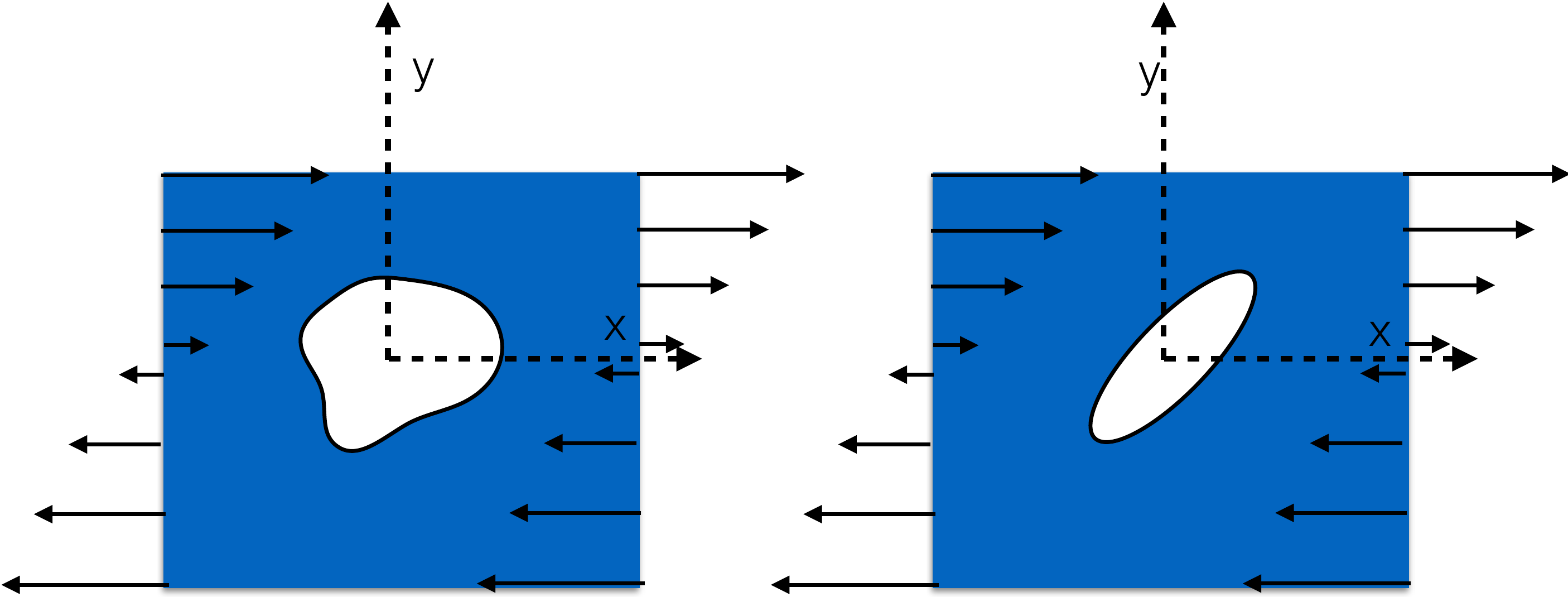}
\caption{Bubble dynamics in an incompressible fluid with odd viscosity placed in a shear flow. Any smooth initial shape leads to an odd viscosity dependent elliptical shape in equilibrium. In the presence of surface tension the ellipse parameters depend both on odd viscosity and surface tension terms.}
     \label{fig:bubblestrain}
\end{figure}
 
\subsection{Bubble dynamics} As a second example of {\it no-stress} boundary condition, we consider the case of an inviscid bubble in  Stokes flow~\cite{richardson1968two, richardson1973two} with odd viscosity. We define the problem by specifying the flow at infinity and monitor the equilibrium shape of the bubble under the specified fluid flow. We will consider a shear flow and hyperbolic flow (also known as straining flow)  as representative cases. The motivation to consider this problem is to capture some observable effects of odd viscosity in the final equilibrium shape of the bubble. The fluid inside the bubble has a negligible viscosity (both dissipative and odd) and is at a constant pressure, which is chosen to be zero without loss of generality. The fluid outside the bubble has a dissipative viscosity $\nu_e$ and odd viscosity $\nu_o$ and is incompressible. Neglecting inertial effects, gravitational or other body forces, the fluid motion is governed by the incompressibility condition and the Stokes equation,
\begin{align}
	\partial_iv_i=0,\quad	\Delta v_i&=\frac{1}{\nu_e}\partial_i \tilde p \,.
 \label{eq:Stokesflow}
\end{align}
We consider the regime with low Reynolds number and drop all non-linear terms in Eq.~\ref{eq:navierStokes}. On the surface of the bubble, we must satisfy two dynamic boundary conditions that balance the forces
\begin{align}
n_jT_{ij}=-\sigma \kappa	n_i,
\label{eq:dbc}
\end{align}
\begin{figure}[tb]
  \centering
\includegraphics[scale=0.32]{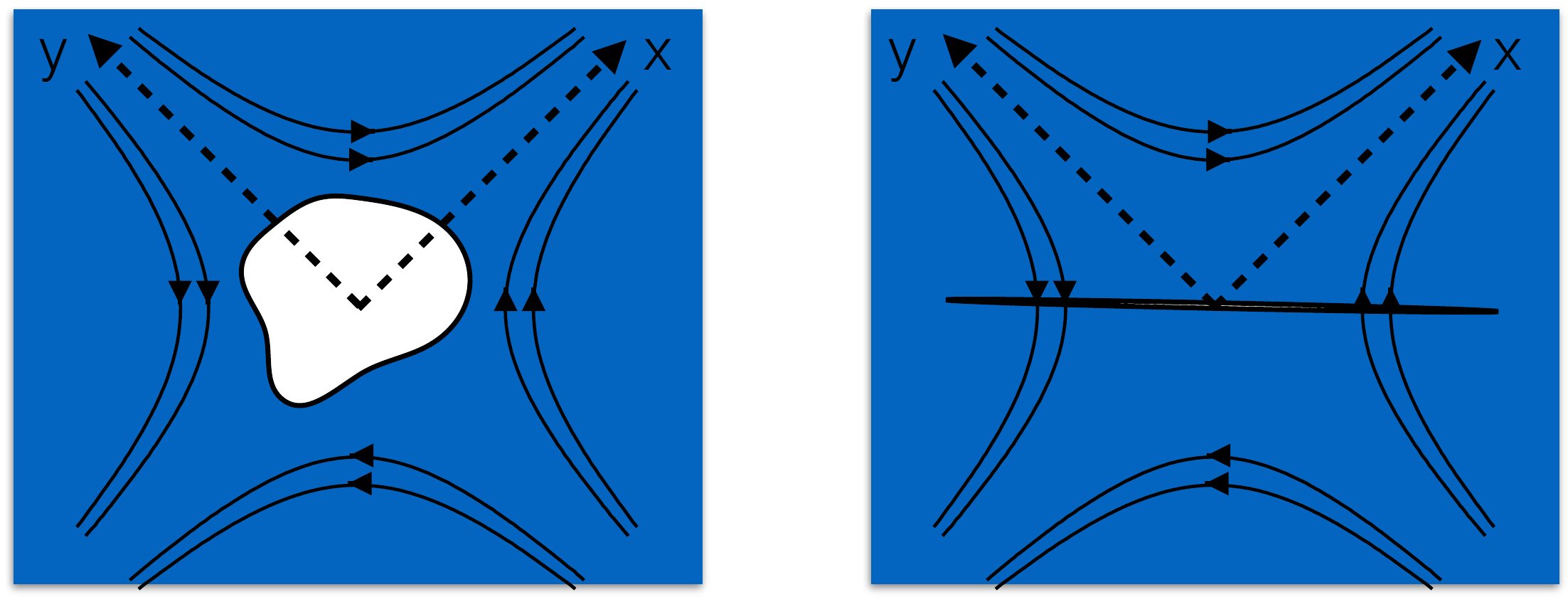}
\caption{Bubble dynamics in an incompressible fluid with odd viscosity placed in a hyperbolic (straining) flow. Any smooth initial shape leads to a slit shape in equilibrium independent of odd viscosity (with no surface tension). Adding surface tension restores the elliptical shape. }
     \label{fig:holes}
\end{figure}
where we have defined $\sigma$ as the surface tension, $\kappa$ as the curvature of the surface, and $n_{x,y}$ as two components of the outward normal unit vector. We also need to match the normal velocity $V_n$ of a point on the bubble surface to the normal component of fluid velocity at that point,
\begin{align}
	v_in_i=V_n\,.
 \label{eq:kbc}	
\end{align}
To completely specify the problem, we must specialize the appropriate boundary conditions at large distances, which can be done by choosing the type of flow. 

\textbf{\textit{Bubble in a shear flow:}} Now we consider a bubble placed within a shear flow such that $(v_x,v_y)\to (k y,0)$ at infinity. Here, $k$ is the shearing parameter or the vorticity at large distances. For this flow, we follow Ref.~\cite{richardson1968two}. Solving the biharmonic equation for the stream function and imposing boundary conditions yields the exact bubble shape in the form of ellipse. The major axis of the ellipse is inclined to the flow direction at an angle $\alpha$ and the deformation parameter of the ellipse $\rho$ (the ratio of the half difference to the half sum of major and minor axes) are given by
\begin{align}
	\rho e^{2i\alpha}=\frac{1}{1-ic} \,.
	\label{eq:ellipse}
\end{align}
Here the real function $c \equiv c(\sigma,\nu_e,\nu_o)$ can be found by matching boundary conditions which fix the pressure at large distances. The details of calculations are given in the appendix. The expression for $c$ is especially simple in the limit of zero surface tension $\sigma=0$ where $c=-\nu_o/\nu_e$. In the absence of odd viscosity $\nu_o$, the ellipse collapses to a slit. In the limit $\nu_e\rightarrow 0$ the bubble shape is near circular, similar to the infinite surface tension limit~\cite{richardson1968two}. Thus the finite odd viscosity stabilizes the ellipse shape by playing a role akin to surface tension. 

\subsection{Bubble in hyperbolic flow and resurrection of flow dependent boundary conditions} We now focus on the scenario when the bubble is placed in a hyperbolic (purely straining) flow $(v_x,v_y)\to C(x,-y)$. For this case, we again obtain elliptical shape defined by Eq.~\ref{eq:ellipse}. However, for this case $c\equiv c(\sigma,\nu_e)$ is independent of the value of odd viscosity. For the $\sigma=0$ limit the basic straining motion transmits only a constant normal force across the real axis, so that the bubble is again a slit along this axis with no perturbation on the flow. The details demonstrating odd viscosity independent flow and boundary conditions for no-stress boundary conditions for purely straining flow are given in the appendix. It appears that the non-vanishing vorticity in the case of shear flow is crucial for the shape of the bubble to depend on odd viscosity. In contrast to the shearing flow, the no-stress conditions for the hyperbolic case on the bubble surface can be effectively reduced to flow dependent boundary conditions, thereby resurrecting statements (I-III). 

\section{Outlook} The presence of the parity breaking odd viscosity term in 2D hydrodynamics leads to many interesting phenomena. Those phenomena have received limited attention barring few exceptions \cite{wiegmann2014anomalous,lapa2014swimming,lucas2014phenomenology,banerjee2017odd}. Here we focus on the case of an incompressible fluid. We show that for no-slip boundary conditions the flow itself does not depend on the value of odd viscosity. We give three exact statements concerning the flow, forces and torques in this case.  Using the non-vanishing torque,  we formulate a possible protocol for measuring odd viscosity and introduce the concept of odd viscostance. The no-stress boundary conditions generically produce the flow dependent on the value of odd viscosity. To this end,  we consider the classical problem of the shape of the bubble in shear and hyperbolic flows and show under what conditions flow is modified by the odd viscosity. A future direction would be to specialize our framework to experimental platforms such as active rotors (see \onlinecite{banerjee2017odd} and references therein).

\textit{Acknowledgements}: We thank D. Banerjee, W. Irvine, G. Monteiro, A. Souslov, V. Vitelli, T. Can and P. Wiegmann for stimulating discussions. A.G.A. acknowledges the financial support of the NSF under grant no. DMR-1606591 and the hospitality of the Kadanoff Center for Theoretical Physics.

\appendix

\section{Two dimensional bubble in Stokes flow with odd viscosity}

Here we consider the problem of a stationary shape of the bubble placed in a two-dimensional slow viscous flow with non-zero odd viscosity. We closely follow the beautiful paper by Richardson~\cite{richardson1968two}. It turns out that the the only technical modification required due to the presence of the odd viscosity is the replacement of the shear viscosity $\nu_e$ by the complex viscosity 
\begin{align}
	\nu=\nu_e+i\nu_o\,.
\end{align}
We refer the reader for all details to \cite{richardson1968two} and present here only results and major steps of the derivation.

\textbf{\textit{Stokes flow:}} 
The fluid inside the bubble has a negligible viscosity (both dissipative and odd) and is at a constant pressure, which is chosen to be zero without loss of generality. The fluid outside the
bubble has a dissipative viscosity $\nu_e$ and odd viscosity $\nu_o$ and is incompressible. Neglecting inertial effects, gravitational or other body forces, the fluid motion is governed by the incompressibility condition and the Stokes equation (\ref{eq:Stokesflow}). We follow \cite{richardson1968two} and introduce complex notations for coordinates $z=x+iy$ and partial derivatives $\p=2(\p_x-i\p_y)$. Then the general solution of (\ref{eq:Stokesflow}) is given in terms of complex velocity $v=v_x-iv_y$ as:
\begin{align}
	v=2i\partial\psi=i(\overline{\phi(z)}+\bar z \phi'(z)+\chi'(z))
 \la{eq:vstokes}
\end{align}
with the stream function $\psi$ and modified pressure $\tilde{p}$ given 
\begin{align}
	\psi(x,y)=\Re(\bar z \phi(z)+\chi(z)),\quad \tilde p/\nu_e+i\omega=-4\phi'(z)	\,.
  \la{eq:psistokes}
\end{align}
Here $\phi(z)$ and $\chi(z)$ are arbitrary functions analytic in $z$ outside the bubble.

\textbf{\textit{No-stress boundary conditions on the surface of the bubble:}} 
On the surface of the bubble, we must satisfy two stress conditions (dynamic boundary conditions) given by Eq.~\ref{eq:dbc} and one kinematic condition Eq.~\ref{eq:kbc}. The stress tensor can be expressed as $T_{ij}=-(\tilde p +\nu_o \omega)\delta_{ij}+\sigma_{ij}$ with the viscous stress from (\ref{sigmaij}). In complex notations
\begin{align}
 	T & \equiv T_{xx}-T_{yy}-2iT_{xy}=4\nu \partial v\,,
 \\
 	\Theta & \equiv T_{xx}+T_{yy} = -2(\tilde{p}+\nu_o\omega) \,.
\end{align}
The stress tensor is conserved $\bar\partial T+\partial\Theta=0$ and we write its components in terms of the complex function $H(z,\bar{z})$ as 
\begin{align}
	T=4i\partial H\,, \quad \Theta =-4i\bar\partial H \,.
 \la{TThetaH}
\end{align} 

The dynamic no-stress boundary conditions (\ref{eq:dbc}) on the surface of the bubble $\Gamma$ can be written in complex notations,
\begin{align}
	NT+\bar{N}\Theta =2\sigma \kappa \bar{N}\,,
 \la{eq:dbc2}
\end{align}
where we have defined the complexified normal vector to $\Gamma$ as $N=n_x+in_y$. If the surface $\Gamma$ is given in parametric form as $z(s)$ with $s$ - the arc length traversed in the clockwise direction,
\begin{align}
	N=i(x_s+iy_s)=iz_s=ie^{i\theta}\,.
 \la{N}
\end{align}
Here $\theta$ is the angle between the tangent and the real positive $x$-axis. The curvature is then defined as $\kappa=-\theta_s=-\bar{z}_{ss}/\bar{N}$ (we use the notation $a_s=\frac{da}{ds}$). We rewrite (\ref{eq:dbc2}) as,
$$
	iz_s T-i\bar{z}_s\Theta=-2\sigma \bar{z}_{ss} \,.
$$
and using (\ref{TThetaH}):
\begin{align}
	\bar z_s \bar\partial H(z,\bar z)+ z_s\partial H(z,\bar z) =-\frac{\sigma}{2}\bar z_{ss}\,.
\end{align}
Integrating over $s$ we obtain the dynamic boundary condition in terms of $H(z,\bar{z})$ valid on $\Gamma$:
\begin{align}
	 H(z,\bar z) =-\frac{\sigma}{2}\bar z_{s}\,.
 \la{eq:dbc3}
\end{align}

Next we consider the kinematic boundary condition (\ref{eq:kbc}) setting the bubble surface velocity to $V_n=0$ (for stationary problem) and in complex notations on $\Gamma$ we have,
\begin{align}
	\Re[v N]=0\,.
 \la{eq:kbc2}
\end{align}
or using (\ref{eq:vstokes},\ref{N}) in terms of the stream function we obtain $-\p_s\psi=0$, i.e.,  taking constant to be zero
\begin{align}
	\psi\Big|_\Gamma =0\,.
 \la{eq:kbc3}
\end{align}

\textbf{\textit{Boundary conditions on bubble for the Stokes flow: }}
Let us now rewrite the boundary conditions (\ref{eq:dbc3},\ref{eq:kbc3}) for the Stokes flow given by (\ref{eq:vstokes},\ref{eq:psistokes}). We obtain:
\begin{align}
	 \nu (\bar z \phi'(z)+\chi'(z))-\overline{\nu\phi(z)} &=-\frac{\sigma}{2}\bar z_{s}\,,
 \label{dbc} \\
	 \Re[\bar z \phi(z)+\chi(z)]	&= 0 \,.
 \label{kbc}
\end{align}
Here we used (\ref{eq:vstokes},\ref{eq:psistokes}) to find
\begin{align}
	H(z,\bar z)=\nu (\bar z \phi'(z)+ \chi'(z)) - \overline{\nu\phi(z)}
\end{align}

Multiplying (\ref{dbc}) by $z_s$, we obtain
\begin{align}
	 \nu \p_s(\bar z \phi(z)+\chi(z)) &=-\frac{\sigma}{2} +2\,\Re[z_s \overline{\nu\phi(z)}]\,.
 \label{dbc4} 
\end{align}
Taking imaginary part of (\ref{dbc4}) and combining with (\ref{kbc}) we obtain 
\begin{align}
	 \bar z \phi(z)+\chi(z)\Big|_\Gamma=0\,
 \label{bc1} 
\end{align}
and taking the real part of (\ref{dbc4}) and using (\ref{bc1})
\begin{align}
	\Re[\bar{z}_s \nu\phi(z)]\Big|_\Gamma=\frac{\sigma}{4}\,.
 \label{bc2} 
\end{align}

Thus the three real boundary conditions (\ref{eq:dbc3},\ref{eq:kbc3}) can be written in a compact form (\ref{bc1},\ref{bc2}) differing from the one written in Ref.~\cite{richardson1968two} only by changing the dissipative viscosity $\nu_e$ to complex viscosity $\nu=\nu_e+i\nu_o$. 
 
\subsection{Two-dimensional bubble in a shear flow} To completely specify the problem, we must specialize the appropriate boundary conditions at large distances. We consider the bubble which is placed in a linear flow.  We consider two examples of linear flow:  shear flow and hyperbolic flow (or purely straining flow). In this section we consider a two-dimensional bubble placed within a shear flow which is specified by the asymptotics at infinity as $\bm{v}=(k y, 0)$. This requires that in the limit $|z|\rightarrow \infty$,
 \begin{align}
 	\phi(z)\sim \frac{1}{4}\left(k-\frac{i\tilde p_{\infty}}{\nu_e}\right)z\,,
	\qquad \chi(z)\sim-\frac{1}{4}k z^2\, ,
 \end{align}
where $\tilde p_{\infty}$ is the yet to be determined pressure at infinity resulting from zero pressure within the bubble. The goal here is to determine the equilibrium shape of the bubble. To achieve this, we use conformal mapping which maps the fluid region to the exterior of the unit circle $\Gamma$ in the $\zeta$-plane, given by $z=w(\zeta)$. $w(\zeta)$ is analytic in the flow domain $|\zeta|\geq 1$ for a smooth bubble outline. We choose a unique mapping requiring $w(\zeta)\sim a \zeta$ as $|\zeta|\rightarrow \infty$ where $a$ is a real constant related to the bubble size. Now we define $\Phi(\zeta)=\phi(w(\zeta)), \quad X(\zeta)=\chi(w(\zeta))$. Both are analytic functions for $|\zeta|\geq 1$, and in the limit $|\zeta|\rightarrow \infty$,
\begin{align}
	 	\Phi(\zeta)\sim \frac{1}{4}\left(k-\frac{i\tilde p_{\infty}}{\nu_e}\right)a\zeta,\quad X(\zeta)\sim -\frac{1}{4}k a^2\zeta^2\,.
\end{align}
The two boundary conditions in Eqs.~\ref{bc1},\ref{bc2} can then be written as,
\begin{align}
	\overline{w(\zeta)}\Phi(\zeta)+X(\zeta)=0, \quad \Re[i \zeta w'(\zeta)\overline{\nu\Phi(\zeta)}]=\frac{\sigma}{4}|w'(\zeta)|\,.
	\label{circle}
\end{align}
 $\zeta$ resides in the exterior of a unit circle and $1/\bar \zeta$ resides in the interior at the inverse point. If $f(\zeta)$ is an analytic function outside of $\Gamma$ then $\overline{f(1/\bar \zeta)}$ is an analytic function of $\zeta$ in the interior. Using the first condition of (\ref{circle}) we analytically continue $w(\zeta)$ to the domain $|\zeta|\leq 1$ as,
\begin{align}
 	w(\zeta)& =-\frac{\overline{X(1/\bar{\zeta})}}{\overline{\Phi(1/\bar{\zeta})}}\,, 
	\;\;\;\text{for}\, |\zeta|\leq 1\,,
  \label{interior} \\
 	w(\zeta) &\sim \frac{a}{1-ic}\frac{1}{\zeta} \,,\;\;\;\;\;\, \text{for}\, |\zeta|\rightarrow 0\,,
\end{align}
 where we have defined $c=-\frac{\tilde p_{\infty}}{k\nu_e}$.
 Replacing $\zeta$ by $1/\bar \zeta$ and taking conjugate in Eqs.~(\ref{interior}), we get,
 \begin{align}
 	\overline{w(1/\bar\zeta)}=-\frac{X(\zeta)}{\Phi(\zeta)}, \quad |\zeta|\geq 1.
 \end{align}
 The second condition of (\ref{circle}) to be satisfied on $|\zeta|=1$ can be written as,
\begin{align}
	\frac{1}{\zeta}\overline{w'( 1/\bar\zeta)} \nu \Phi(\zeta)
	-\zeta w'(\zeta)\overline{\nu \Phi(\zeta)}	
	=\frac{i\sigma}{2}[w'(\zeta)\overline{w'( 1/\bar\zeta)}]^{1/2}\,.
 \nonumber
\end{align}
We analytically continue this condition to the whole plane,
\begin{align}
	\nu\frac{\Phi(\zeta)}{\zeta w'(\zeta)}
	-\bar\nu \zeta\frac{\overline{\Phi(1/\bar\zeta)}}{\overline{w'( 1/\bar\zeta)}}
	=\frac{i\sigma}{2}\frac{1}{[w'(\zeta)\overline{w'( 1/\bar\zeta)}]^{1/2}}\,.
 \la{ancon}
\end{align}
The first term on the left-hand side is analytic and single-valued in $|\zeta|\geq 1$ and the second term in $|\zeta|\leq 1$. The condition (\ref{ancon}) is identical to the one obtained in Ref.~\cite{richardson1968two} with the only change that $\nu$ is complex. For real $\nu$ as in \cite{richardson1968two} the limit $\sigma\to0$ is somewhat peculiar. For complex $\nu$ we can just put $\sigma=0$ in (\ref{ancon}) in the absence of the surface tension. We consider two cases.
 
\emph{Case I: $\sigma=0$:} In this case the functional relationship (\ref{ancon}) becomes,
 \begin{align}
 	\nu\frac{\Phi(\zeta)}{\zeta w'(\zeta)}
	=\bar\nu \zeta \frac{\overline{\Phi(1/\bar\zeta)}}{\overline{w'(1/\bar\zeta)}}\,.
 \la{anconS0}
\end{align}
The left-hand side is analytic and single-valued in $|\zeta|\geq 1$ and the right hand side $|\zeta|\leq 1$. Therefore, both sides should be equal to the same constant. We know the limiting cases in each limit:
 \begin{align}
 	\nu\frac{\Phi(\zeta)}{\zeta w'(\zeta)}
	& \sim \nu\frac{k}{4}(1+ic)\,, 
	\quad |\zeta|\rightarrow \infty\,,
 \\
 	 \bar\nu \zeta \frac{\overline{\Phi(1/\bar\zeta)}}{\overline{w'( 1/\bar\zeta)}}
	 & \sim \bar \nu\frac{k}{4}(1-ic), \quad |\zeta|\rightarrow 0\,.
\end{align}
Thus we have for $c$,
\begin{align}
 	\nu\frac{k}{4}(1+ic)=\bar\nu\frac{k}{4}(1-ic) 
	\quad\implies
 	c=-\frac{\nu_o}{\nu_e}\,.
\end{align}
The general form for $w(\zeta)$ can be written as \cite{richardson1968two},
\begin{align}
 	w(\zeta)=a(\zeta+\gamma^2/\zeta), 
 \label{eq:map}
\end{align}
 where $\gamma^2=\frac{1}{1-ic}$. This mapping with $\gamma^2=\rho e^{2i\alpha}$ gives an ellipse with its major axis inclined at an angle $\alpha$ to the flow direction and the deformation parameter of the ellipse $\rho$ (the ratio of the half difference to the half sum of major and minor axes).  The ellipse parameters are given by:
\begin{align}
 	\tan2\alpha=-\frac{\nu_o}{\nu_e}, \qquad \rho=\frac{1}{1+\nu_o^2/\nu_e^2}\,.	
\end{align}
 
  \emph{Case II: $\sigma\ne 0$:} In the presence of surface tension $\sigma$, the ellipse parameters depend on $\sigma$ via a transcendental equation for $c$~\cite{richardson1968two},
  \begin{align}
  	c=-\frac{\nu_o}{\nu_e}+\frac{2 \sigma}{\pi a \nu_e k}	K[(1+c^2)^{-1/2}].
  \end{align}
Here $K(m)=\int_0^1\frac{ds}{\sqrt{(1-s^2)(1-m^2s^2)}}$ is the complete elliptic integral of the first kind.

\subsection{Two dimensional bubble in a hyperbolic flow} We now consider the bubble placed in a hyperbolic flow (a pure straining flow) $\bm{v}=(Cx,-Cy)$, where $C>0$. Such a flow is similar to a rigid rotation superposed on top of a simple shear of strength $k=2C$. For this flow we note that,
\begin{align}
 	\phi(z)\sim \frac{i\tilde p_{\infty}}{4\nu_e}z\,,
	\quad \chi(z)
	\sim-\frac{iC}{2}z^2, \quad |z|\rightarrow \infty\,,
 \end{align}
where $\tilde p_{\infty}$ is yet to be determined pressure at $\infty$.  Proceeding as in the previous section we map to the $\zeta$-plane where,
\begin{align}
 	\Phi(\zeta)\sim \frac{i\tilde p_{\infty}}{4\nu_e}a\zeta,\quad X(\zeta)
	=-\frac{iC}{2}a^2\zeta^2 \,, 
	\quad |\zeta|\rightarrow \infty\,.
 \end{align}
The analytic continuation of $w(\zeta)$ into the interior of $|\zeta|=1$ now yields,
\begin{align}
	w(\zeta)=-\frac{2aC \nu_e}{\tilde p_{\infty} \zeta}\,,
	\quad |\zeta|\rightarrow 0\,.	
\end{align}
The key equation that determines the steady state shape of the bubble in the limit of $\sigma=0$ is again given by (\ref{anconS0}).
The left-hand side of (\ref{anconS0}) is analytic and single-valued in $|\zeta|\geq 1$ and the right hand side $|\zeta|\leq 1$. In the limits:
 \begin{align}
 	\nu\frac{\Phi(\zeta)}{\zeta w'(\zeta)}
	&=-i\nu\frac{\tilde p_{\infty}}{4\nu_e}\,, \;\quad |\zeta|\rightarrow \infty \,,
 \la{as1} \\
 	\bar\nu \zeta \frac{\overline{\Phi(1/\bar\zeta)}}{\overline{w'( 1/\bar\zeta)}}
	&=i\bar\nu\frac{\tilde p_{\infty}}{4\nu_e}\,, \qquad |\zeta|\rightarrow 0 \,.
 \la{as2}
 \end{align}
The only way to satisfy (\ref{anconS0}) is to have $\tilde p_{\infty}=0$. Notice that the odd viscosity does not enter this solution and the bubble is simply a slit along the real axis. In the presence of surface tension $\sigma\ne 0$ the boundary condition yields the equation of ellipse given again by  Eq.~\ref{eq:map} with $\gamma^2=-2C\nu_e/\tilde p_{\infty}$. $\gamma^2$ can be found by solving the transcendental equation \cite{richardson1968two},
\begin{align}
 	\frac{1}{\gamma^2}=\frac{\sigma}{\pi a \nu_e C}K(\gamma^2)\,.	
 \la{hypergamma}
\end{align}
The above equation is identical to the results of Ref.~\cite{richardson1968two} with odd viscosity not entering (\ref{hypergamma}). This is to be expected after the observation that putting $\tilde p_{\infty}=0$ into (\ref{as1},\ref{as2}) removes the dependence on odd viscosity from boundary conditions. Therefore, this case is similar to flow dependent boundary conditions independent of odd viscosity and our exact statements apply to this case.

\bibliographystyle{my-refs}
\bibliography{oddviscosity-bibliography.bib}

\end{document}